\documentclass[10pt,preprint]{aastex}

\shorttitle{Solar Microwave QPP with Fine Structures}
\shortauthors{Tan et al.}

\begin{document}


\title{Microwave Quasi-periodic Pulsation with Millisecond Bursts in A Solar Flare on 2011 August 9}
\author{Baolin Tan\altaffilmark{1}, \& Chengming Tan\altaffilmark{1}}
\affil{$^{1}$Key Laboratory of Solar Activity, National
Astronomical Observatories of the Chinese Academy of Sciences,
Datun Road A20, Chaoyang District, Beijing 100012, China. \\
Email: bltan@nao.cas.cn}

\begin{abstract}

An peculiar microwave quasi-periodic pulsation (QPP) accompanying
with a hard X-ray (HXR) QPP of about 20 s duration occurred just
before the maximum of an X6.9 solar flare on 2011 August 9. The
most interesting is that the microwave QPP is consisting of
millisecond timescale superfine structures. Each microwave QPP
pulse is made up of clusters of millisecond spike bursts or narrow
band type III bursts. There are three different frequency drift
rates: global frequency drift rate of microwave QPP pulse group,
frequency drift rate of microwave QPP pulse, and frequency drift
rate of individual millisecond spikes or type III bursts. The
physical analysis indicates that the energetic electrons
accelerating from a large-scale highly dynamic magnetic
reconnecting current sheet above the flaring loop propagate
downwards, impact on the flaring plasma loop, and produce HXR
bursts. The tearing-mode (TM) oscillations in the current sheet
modulate HXR emission and generate HXR QPP; the energetic
electrons propagating downwards produce Langmuir turbulence and
plasma waves, result in plasma emission. The modulation of TM
oscillation on the plasma emission in the current-carrying plasma
loop may generate microwave QPP. The TM instability produces
magnetic islands in the loop. Each X-point will be a small
reconnection site and accelerate the ambient electrons. These
accelerated electrons impact on the ambient plasma and trigger the
millisecond spike clusters or the group of type III bursts.
Possibly each millisecond spike burst or type III burst is one of
the elementary burst (EB). Large numbers of such EB clusters form
an intense flaring microwave burst.

\end{abstract}

\keywords{Sun: flares --- Sun: fine structure --- Sun: microwave
radiation}

\section{Introduction}

On 2011 August 9, a most powerful X6.9 solar flare took place in
active region NOAA 11263, near west limb on the solar disk (left
panel of Figure 1, observed at EUV wavelength of 171\AA~ by
Atmospheric Imaging Assembly on Solar Dynamics Observatory
(AIA/SDO)). The X6.9 flare event starts at 08:00 UT, reaches to
the maximum at 08:04 UT, and ends at 08:14 UT (the top-right panel
in Figure 1). It is the largest one in the current solar Schwabe
cycle, resulting in a coronal mass ejection. Accompanied with this
flare, an strongly microwave burst (right panels of Figure 1) was
observed at a frequency of 2.60 - 3.80 GHz by the Chinese Solar
Broadband Radio Spectrometer in Huairou (SBRS/Huairou). The
microwave burst starts at 08:01 UT, ends at 08:07 UT, lasts for
only about 6 minutes. As a comparison, we know that the microwave
burst associated with other X-class flare event always has a
long-duration of several tens of minutes or 1 - 2 hours, for
example, the X3.4 flare event on 2006 December 13 has a duration
of 110 minutes (Tan et al. 2010). In the peculiar short duration
of microwave burst in the X6.9 flare event, the most prominent
feature is the quasi-periodic pulsation (QPP) and its accompanied
superfine structures of millisecond timescales. The main task of
this work is to investigate the peculiar features of the microwave
QPP in detail and its related physical processes.

It is well known that the flare-related QPP, especially at the
microwave frequency range, can be a valuable diagnostic tool for
in situ conditions in the flaring source region, and the
investigations can obtain unique insight into the coronal plasma
dynamic processes. For example, the duration of a QPP can be a
measure of the plasma density inhomogeneity in the source region,
and the periodicity can be a measure of the diameter or width of
the inhomogeneity, and so on (Roberts et al. 1984). Many people
studied the features and physical mechanism from observations and
theoretical models (Young et al. 1961, Gotwols 1972, Fu et al.
1990, Kliem et al. 2000, Nakariakov et al. 2003, Tan et al. 2007,
etc.). Aschwanden (1987) presented an extensive review of QPP
models and classified them into three groups: (1) MHD oscillations
modulate the microwave emissivity with standing or propagating
waves (Roberts et al. 1984, Nakariakov \& Melnikov 2009), (2)
periodic self-organizing systems of plasma instability, and (3)
the periodic particle accelerations. Different kind of model can
explain different timescale QPP. Recently, a series of microwave
QPPs with multi-timescale and forming a broad hierarchy are also
reported and proposed that different timescales of QPP may have
different generation mechanisms. The broad hierarchy of timescales
of QPP that occurred in a same flare event may imply that there is
a multi-scale hierarchy of sizes of the magnetic configurations
and the timescales of the dynamic magnetic reconnection processes
in the flaring region (Tan et al. 2010).

On the spectrogram of broadband microwave observations, the QPP
behaves as a train of approximated vertical bright stripes with
almost equidistance. Each bright stripe is a QPP pulse. Generally,
the previous works regarded the pulse as a relatively isolated
element. So far, there is very rare work to investigate if there
are some further superfine structures in each QPP pulse. Chernov
et al. (2008) reported a flare event of which some pulses of the
weakly rapid microwave QPP were completely consisting of several
small-scale narrow band drifting millisecond fibers, and the
frequency drift rate of the fibers were in the range of from -160
MHz s$^{-1}$ to -270 MHz s$^{-1}$. In this work, we find that each
individual pulse of the rapid microwave QPP is made up of clusters
of millisecond spike bursts or narrow band type III bursts with
frequency drift rates of about 20 GHz s$^{-1}$. We will introduce
the peculiar features of the superfine structures in the QPP in
Section 2, and make a physical analysis in Section 3. Finally,
some conclusions are obtained in Section 4.

\section{Observations and Data Analysis}

\subsection{Observational Data and Analysis Method}

In this work, we mainly apply the microwave observations obtained
from SBRS/Huairou to investigate the detailed peculiar features of
the microwave QPP. SBRS/Huairou is an advanced solar radio
telescope with super high cadence, broad frequency bandwidth, and
high frequency resolution, which can distinguish the super fine
structures of microwave bursts from the spectrogram (Fu et al.
1995, 2004, Yan et al. 2002). It includes 3 parts: 1.10 - 2.06 GHz
(with the antenna diameter of 7.0 m, cadence of 5 ms, frequency
resolution of 4 MHz), 2.60 - 3.80 GHz (with the antenna diameter
of 3.2 m, cadence of 8 ms, frequency resolution of 10 MHz), and
5.20 - 7.60 GHz (share the same antenna of the second part,
cadence of 5 ms, frequency resolution of 20 MHz). The antenna
points to the center of solar disk automatically controlled by a
computer. The spectrometer can receive the total flux of solar
radio emission with dual circular polarization (left- and
right-handed circular polarization), and the dynamic range is 10
dB above quiet solar background emission. And the observation
sensitivity is: $S/S_{\bigodot}\leq 2\%$, here $S_{\bigodot}$ is
the standard flux value of the quiet Sun. From the Solar
Geophysical Data (SGD) we can obtain the data at frequencies of
1415 MHz, 2695 MHz, 2800 MHz, and 4995 MHz, and make the
calibration of the observational data followed the method reported
by Tanaka et al. (1973). As for the strong burst, the receiver may
work beyond its linear range and a nonlinear calibration method
will be used instead (Yan et al. 2002).

Similar to other congeneric instruments, such as Phoenix (100 -
4000 MHz, Benz et al. 1991), Ond\'rejov (800 - 4500 MHz, Jiricka
et al. 1993) and Brazilian Broadband Spectrometer (BBS, 200 - 2500
MHz, Sawant et al. 2001), SBRS/Huairou has no spatial resolution.
However, as the Sun is a strong radio emission source, a great
deal of works (e.g. Dulk 1985, etc.) show that the microwave
bursts received by spectrometers are always coming from the solar
active region when the antenna points to the Sun.

In order to make the QPP clearer and more reliable, we adopt the
analysis methods which is described in detail by Tan et al.
(2010). With this method, we can obtain the related parameters of
the microwave QPP easily, such as the period, duration,
polarization degree, the global frequency drifting rates (GFDR),
and the single-pulse frequency drifting rates (SPFDR), etc. These
parameters are explained by Tan (2008). Because of the close
relationships between the hard X-ray (HXR) emission and the
microwave bursts (Dennis 1988), we also adopt the HXR observations
obtained by the Reuven Ramaty High Energy Solar Spectroscopic
Imaging (RHESSI) to make a complementary comparison.

\subsection{Observational Results}

During the X6.9 flare, SBRS/Huairou obtained high-quality
observations at a frequency of 2.60 - 3.80 GHz. The right panels
of Figure 1 present the profiles of the microwave flux at a
frequency of 2.80 GHz with left- and right-handed circular
polarization around the flaring event. As a comparison, the
profiles of GOES soft X-ray (SXR) intensities at wavelengths of 1
- 8 \AA (GOES8) and 0.5 - 4 \AA (GOES4) are also plotted in Figure
1. Here, we find that the main part of the microwave burst
occurred in the flare rising and peak phase, and decay rapidly
after the flare peak. The microwave intensity profile has two
obvious enhancements, the first one is at about 08:02:04 UT, the
second one is at about 08:03:56 UT. A obvious QPP occurred in
08:03:09 - 08:03:30 UT, just in the midst of the two enhancements
(marked with thick bidirectional arrow in the right panel of
Figure 1), very closed to the maximum of the SXR GOES flare (08:04
UT).

\begin{figure}   
\begin{center}
 \includegraphics[height=6.0 cm]{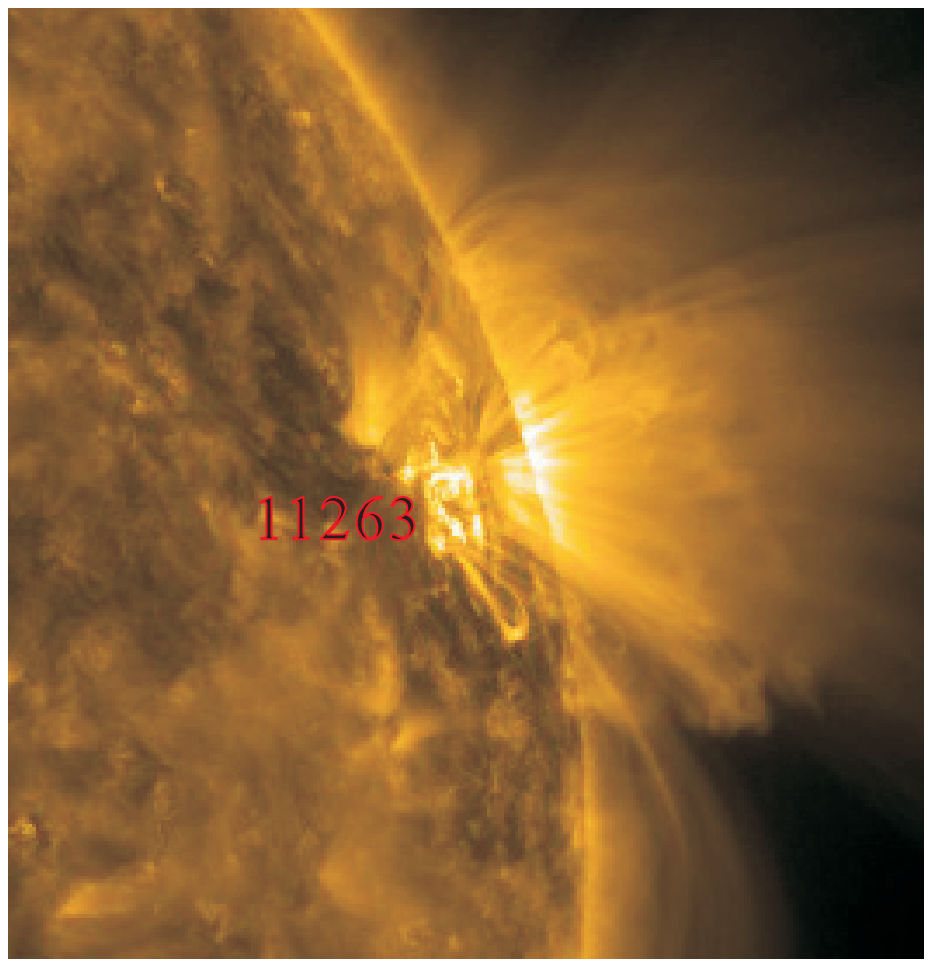}
 \includegraphics[height=6.0 cm]{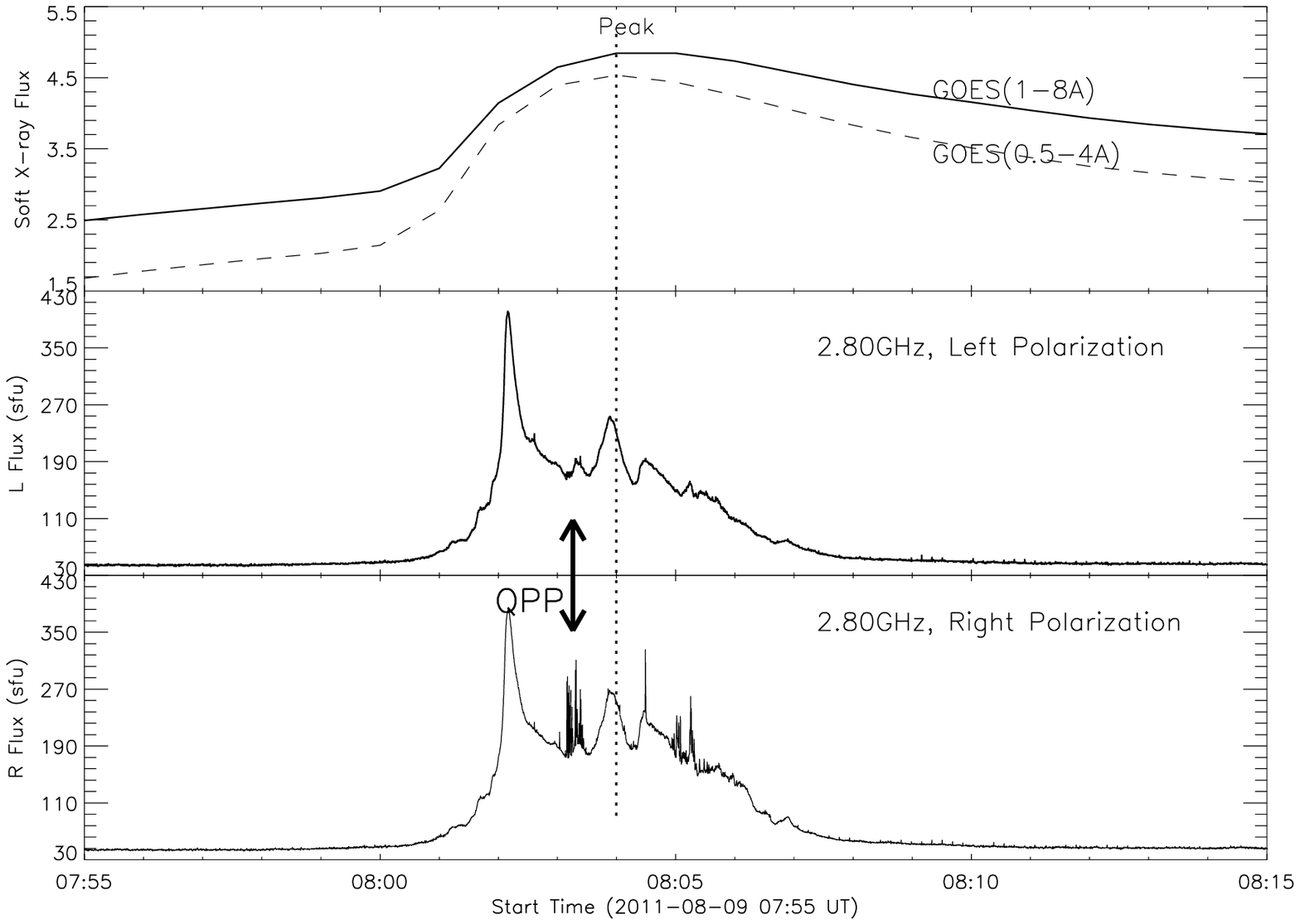}
  \caption{Left panel shows the topology and position of the flaring active region on the solar disk observed by SDO/AIA at
  171 \AA. Right panels show the profiles of GOES soft X-ray (upper), and microwave emission (middle and lower panel) at 2.80 GHz in left- and right-handed polarization
  observed by the Chinese Solar Broadband Radio Spectrometer (SBRS/Huairou). The thick bidirectional arrow indicates the quasi-periodic pulsation (QPP).}
\end{center}
\end{figure}

The careful scrutinizing of the RHESSI HXR observations displays
that HXR emission are also existing QPP features at energy of 12 -
100 keV. The wavelet analysis indicates the evidence of HXR QPP is
most obvious at energy of 25 - 50 keV. Figure 2 presents the light
curve of HXR at energy of 25 - 50 keV during the microwave QPP
occurrence, which shows that the intensity of HXR emission also
has an obvious feature of QPP, and the average period is about
2.46 s, which belongs to short period pulsation (SPP).

\begin{figure}   
\begin{center}
 \includegraphics[width=8.0 cm]{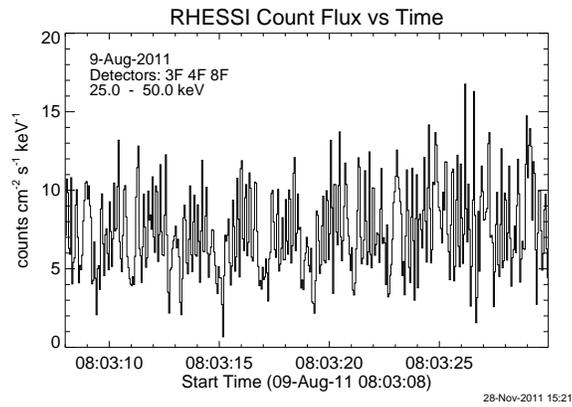}
  \caption{The light curve of hard X-ray at energy of 25 - 50 keV observed by RHESSI during the QPP occurrence, which also behaves the feature of
quasi-periodic pulsations.}
\end{center}
\end{figure}

Figure 3 presents the microwave QPP observed at a frequency of
2.60 - 3.80 GHz by SBRS/Huairou. The top-left panels show the
spectrogram of the QPP at left- and right-handed circular
polarization, and the QPP behaves as a series of bright pulses,
which indicates that the QPP can be divided into three different
pulse groups: the first pulse group begins at 08:03:09 UT, ends at
08:03:17 UT, lasts for 8 s (marked as group A); the second begins
at 08:03:18, ends at 08:03:22 UT, lasts for 4 s (marked as group
B); and the third begins at 08:03:23, ends at 08:03:29 UT, and
lasts for 6 s (marked as group C). The frequency bandwidth of each
pulses is in the range of 400 - 1100 MHz. The comparison between
the left- and right-handed circular polarizations show that the
QPP is strongly right-handed circular polarized with the
polarization degree of about 70\%. The measurement of the time
intervals between each adjacent vertical bright stripe indicate
that the period of the QPP are about 0.70s, 0.86s, and 0.42s at
the first, second, and third pulse group, respectively. All of
them are belong to broad bandwidth very short-period pulsation
(VSP) (Wang \& Xie 2000, Tan et al. 2007). The comparison between
the microwave and HXR, we find that the period of the microwave
QPP is much shorter than that of the HXR QPP. The top-right panels
of Figure 3 show the profiles of the flux intensities of the
microwave QPP at left- and right-handed circular polarization at a
frequency of 2.84 GHz (above and middle panel). It indicates that
the emission enhancement at each QPP pulse respect to the
background emission is about 100 - 250 sfu, which demonstrates
that the QPP is very obvious and strong.

The arranged patterns of microwave QPP pulse groups indicate that
there are frequency drifts in each pulse group (GFDR). In order to
obtain the frequency drift rate, we define the central frequency
of each pulse at its central time as follows:

\begin{equation}
f_{cp}=\frac{\sum(f_{i}\times F_{i})}{\sum F_{i}}
\end{equation}

Here, the central time of the pulse is when the flux intensity
approaches to the maximum around the pulse. $f_{i}$ is the
frequency, and $F_{i}$ is the corresponding flux intensity which
subtract the background emission around the occurrence of the
microwave QPP. The below of the top-right panel of Figure 3
presents the distribution of the central frequency ($f_{cp}$) at
each QPP pulse. The dotted lines are fitted by linear square-least
methods. In QPP group A, the central frequency of QPP pulses can
be fitted by a single line, the slopes of the fitted lines gives
the average GFDR as: 53.3 MHz s$^{-1}$. In QPP group B, the
central frequency of QPP pulses also can be fitted by a single
line with GFDR as -78.9 MHz s$^{-1}$. In QPP group C, the central
frequency of QPP pulses decreases slowly at the first half with
GFDR of -98.2 MHz s$^{-1}$, and then turns to increase slowly with
GFDR of 129.5 MHz s$^{-1}$ at the second half. It seems that there
are some dynamic processes occurred in the emission source region.

\begin{figure}   
\begin{center}
 \includegraphics[width=7.5 cm, height=8.8 cm]{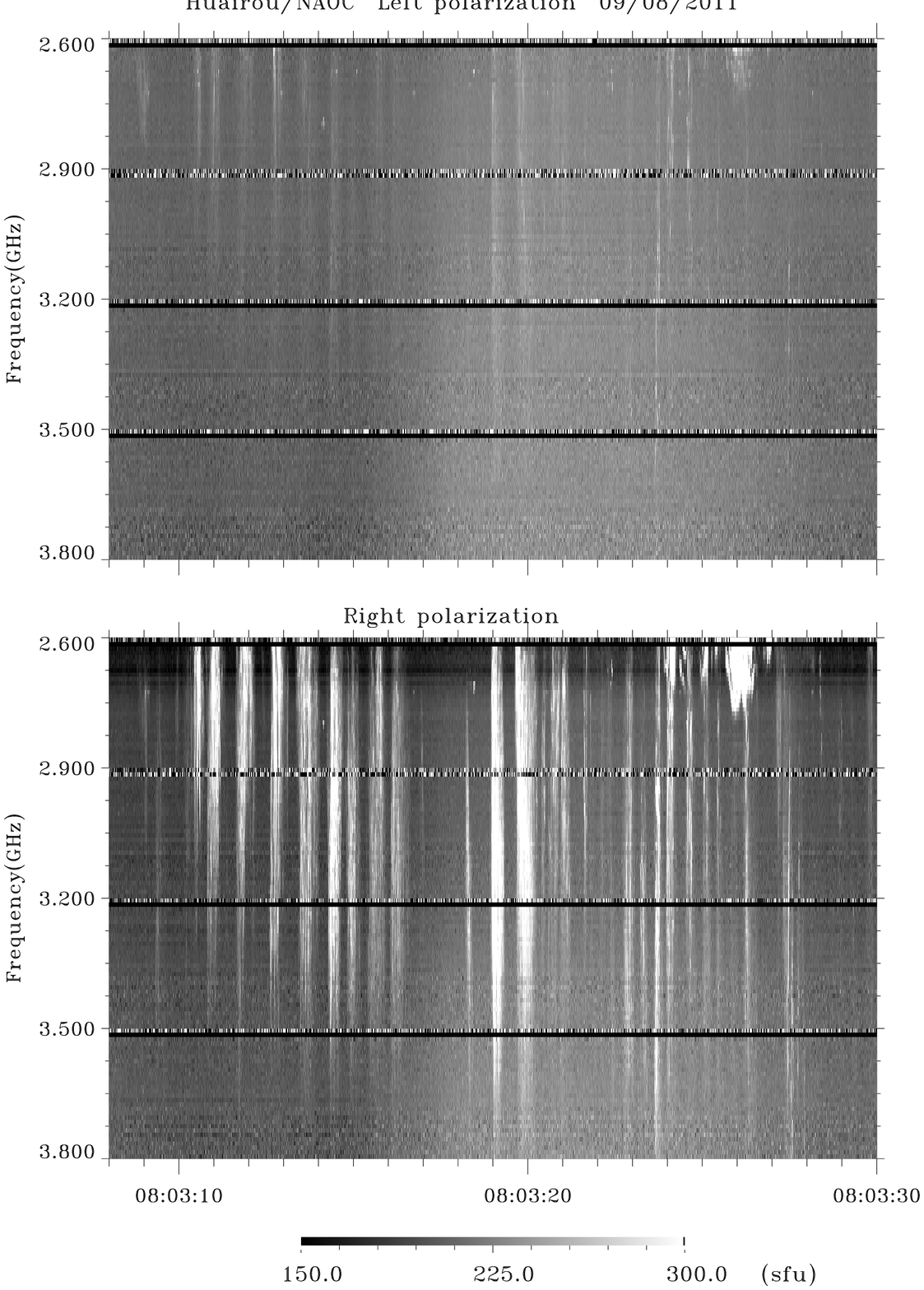}
 \includegraphics[width=8.0 cm, height=8.1 cm]{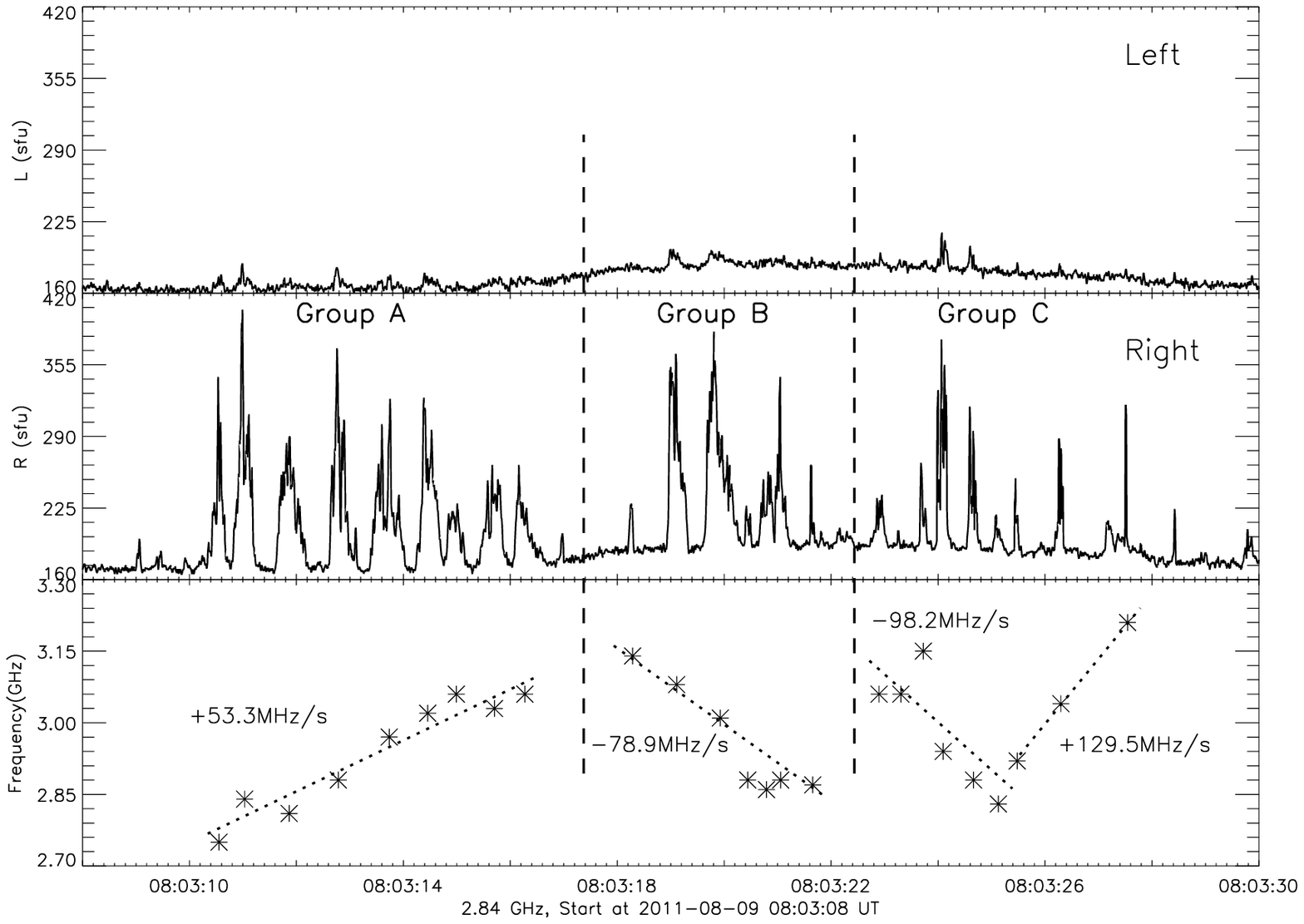}
  \includegraphics[width=5.4 cm]{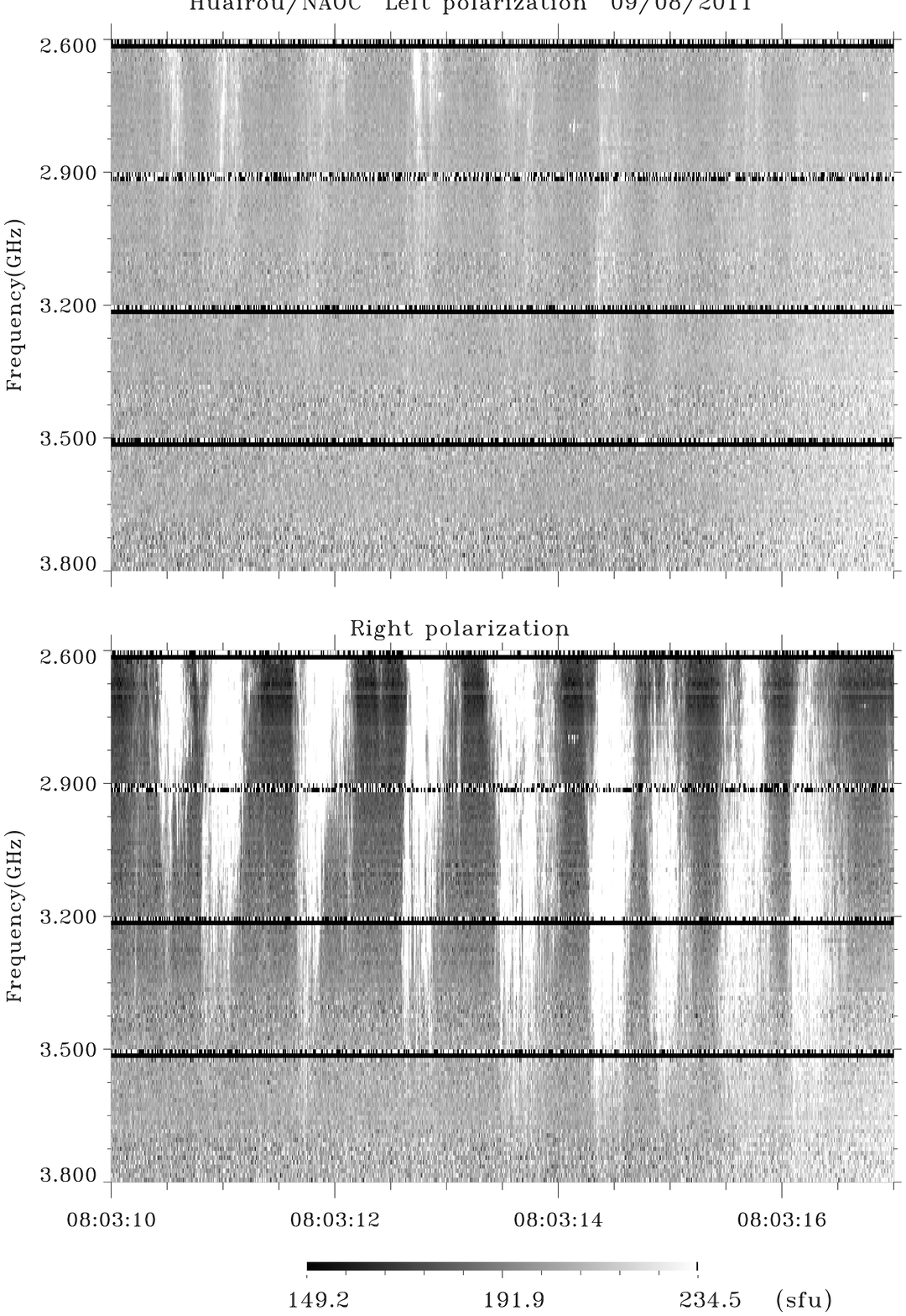}
  \includegraphics[width=5.4 cm]{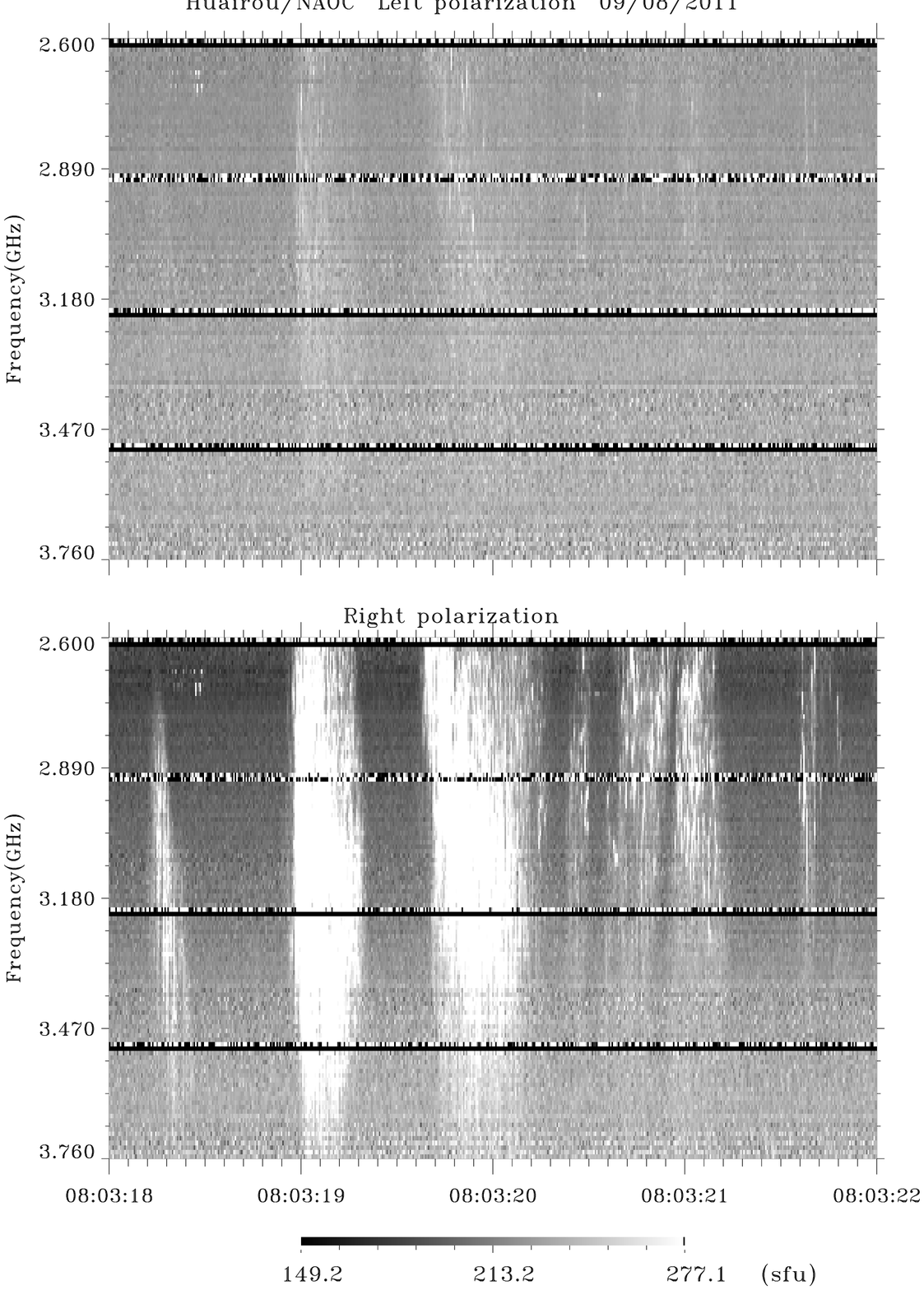}
  \includegraphics[width=5.4 cm]{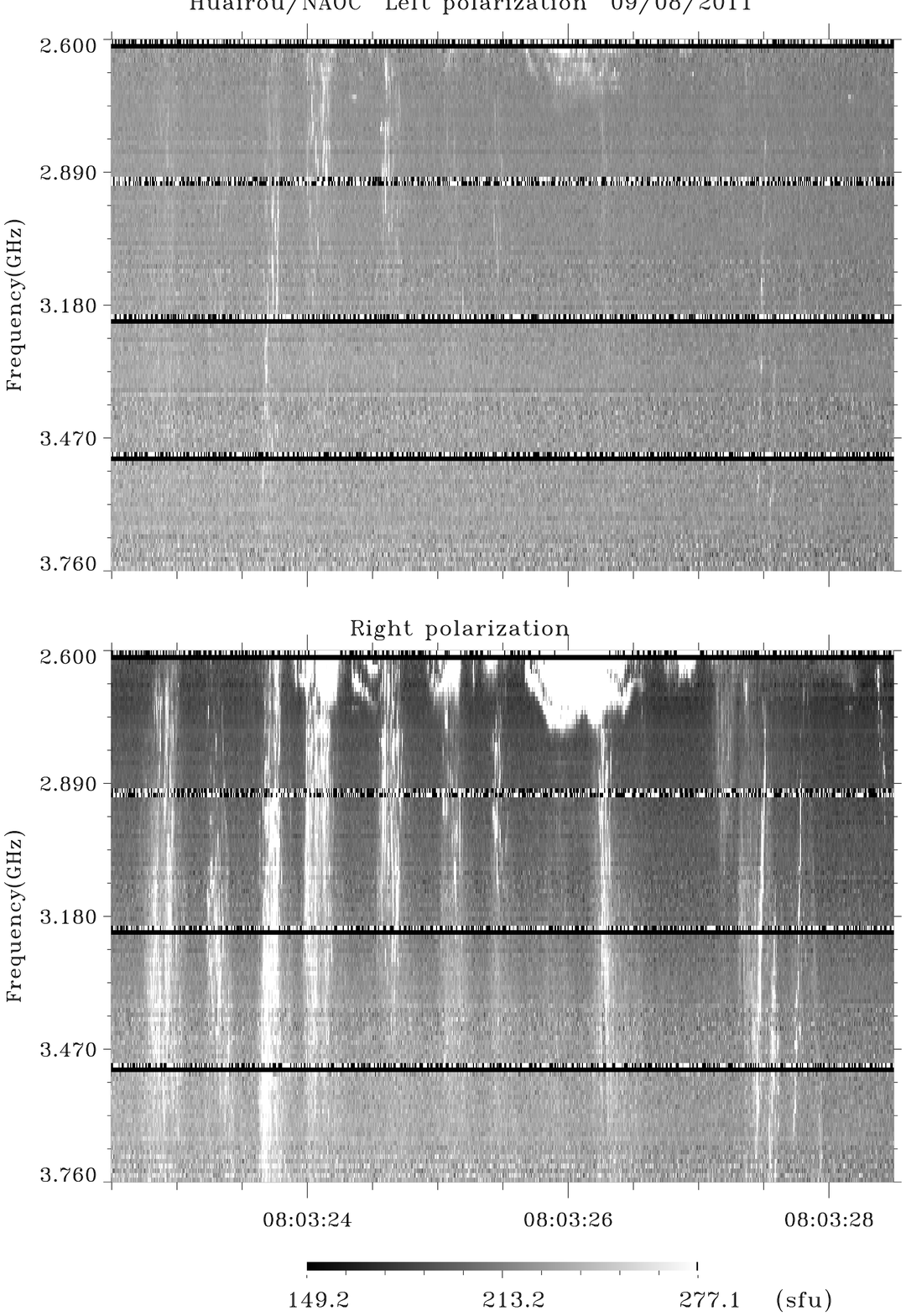}
  \caption{The top-left panels are the spectrogram at left- and right-handed circular polarization of the microwave quasi-periodic pulsation (QPP)
  observed at 2.60 - 3.80 GHz by the Chinese Solar Broadband Radio Spectrometer (SBRS/Huairou). The top-right panels are the emission flux
  profiles at left- and right-handed circular polarizations (above and middle) of the QPP at frequency of 2.60 - 3.80 GHz, and the central
  frequency distribution of QPP pulses (below). The bottom panels are the expanded spectrograms of microwave QPP pulse group A (left),
  B (middle), and C(right), respectively, which can present the details of the superfine structures of QPP pulses. }
\end{center}
\end{figure}

The bottom panels of Figure 3 present the expanded spectrograms of
QPP pulse group A (left), B (middle), and C(right), at left- and
right-handed circular polarization, respectively, which can
present the details of the superfine structures on each QPP pulse.
With a careful scrutinizing, we find that the frequency drifts are
also occurred at each QPP pulse (SPFDR). We apply the similar
method as described in Tan et al. (2010) to determine the SPFDR.
The top-left panel of Figure 4 presents examples of the method,
the SPFDR can be calculated by the slope rate of the fitted lines.
The top-right panel of Figure 4 plots the calculated results of
the SPFDR in the whole QPP. Here, we find that all of the SPFDRs
of QPP pulse group A are negative, and their absolute values are
in the range of 3.51 - 13.84 GHz s$^{-1}$, the average value is
7.73 GHz s$^{-1}$. However, the SPFDRs in QPP pulse group B and C
are rapidly changed from negative to positive, and the absolute
values are from 4.38 GHz s$^{-1}$ to 14.61 GHz s$^{-1}$ in group B
and 2.91 GHz s$^{-1}$ to 26.25 GHz s$^{-1}$ in group C, the
average value is about 7.78 GHz s$^{-1}$ in group B and 15.88 GHz
s$^{-1}$ in group C. Additionally, almost all of the SPFDRs are at
least two orders higher than that of GFDR in the QPP pulse groups.

\begin{figure}         
\begin{center}
  \includegraphics[width=7.5 cm]{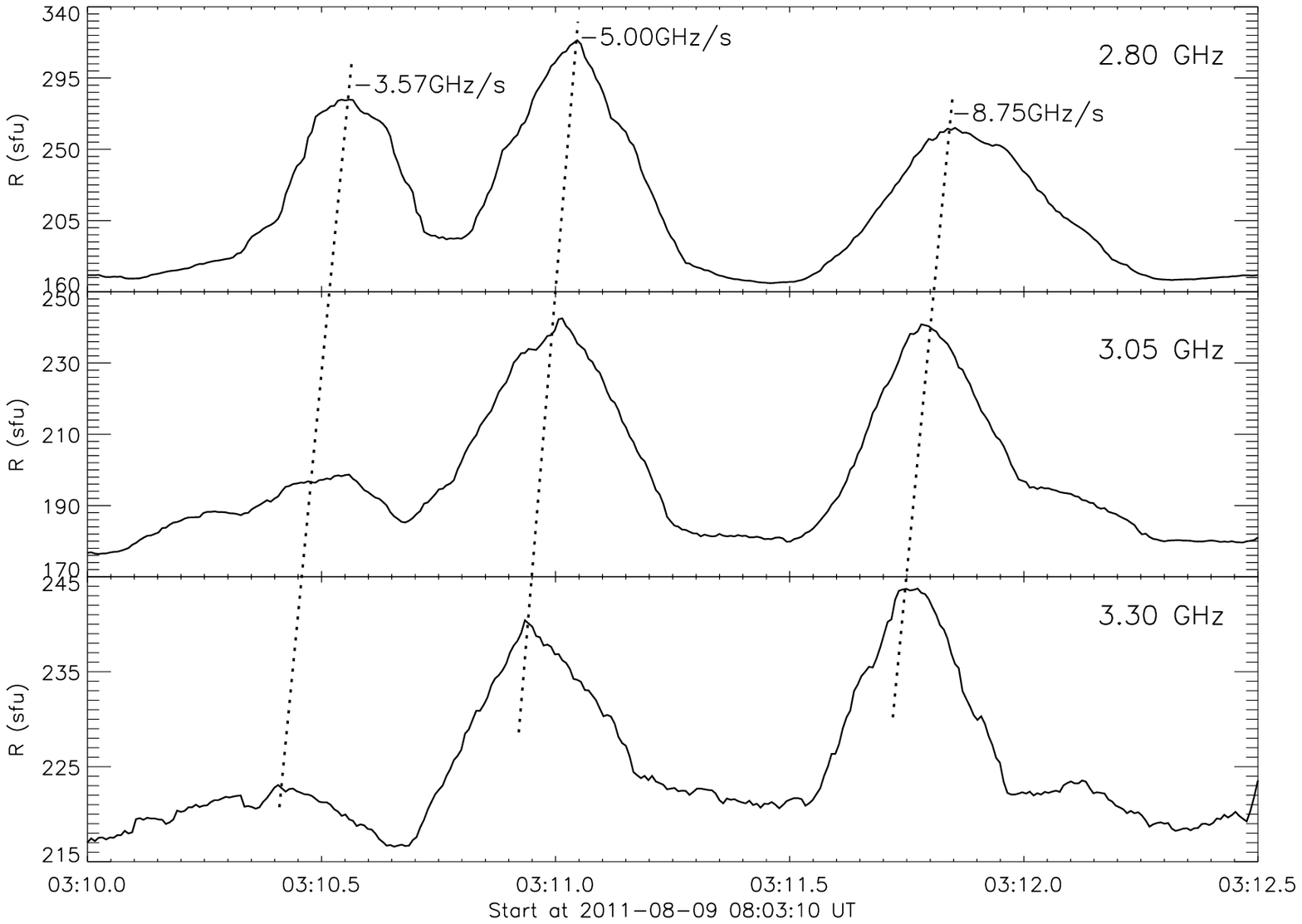}
  \includegraphics[width=7.5 cm]{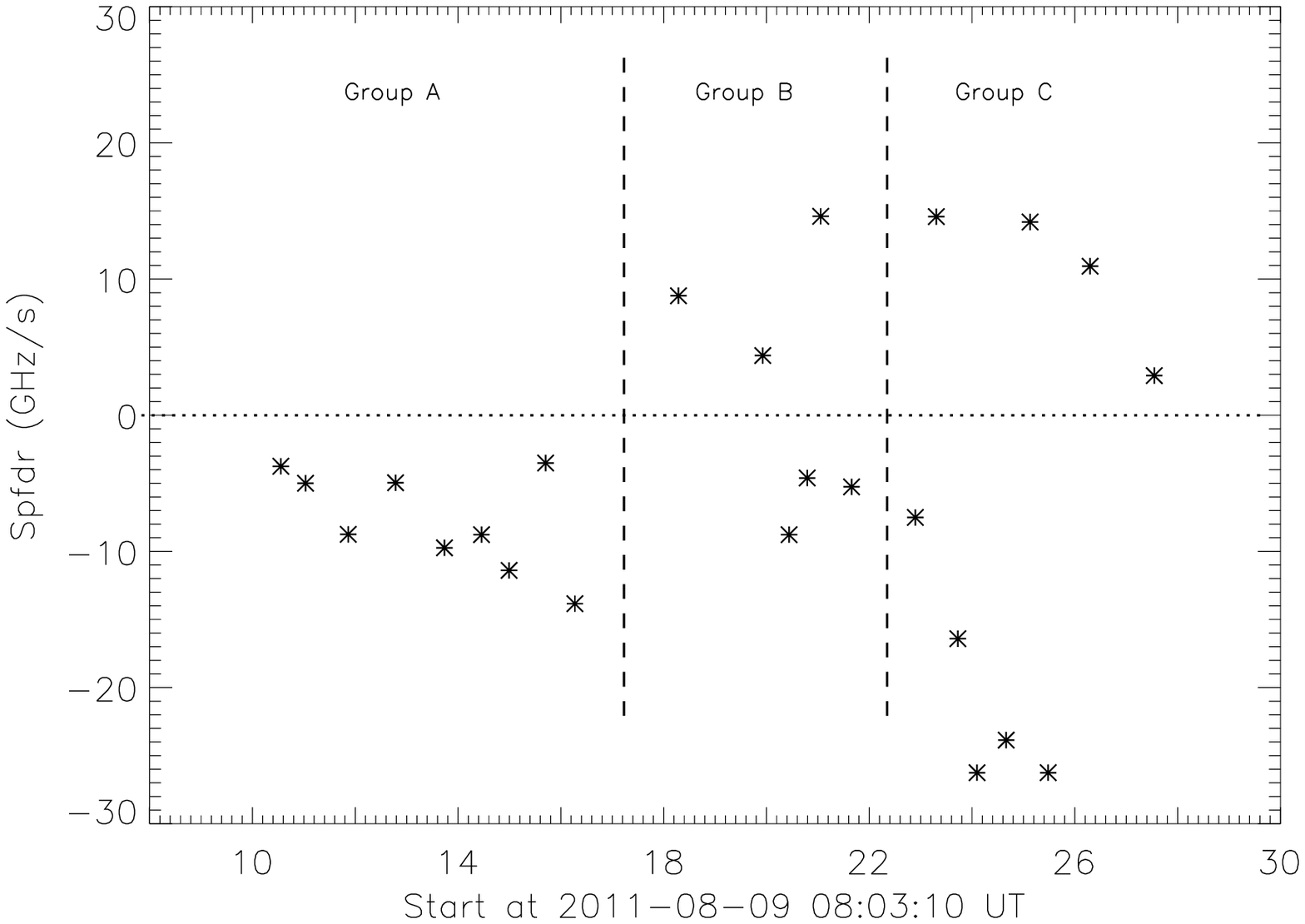}
  \includegraphics[width=7.5 cm]{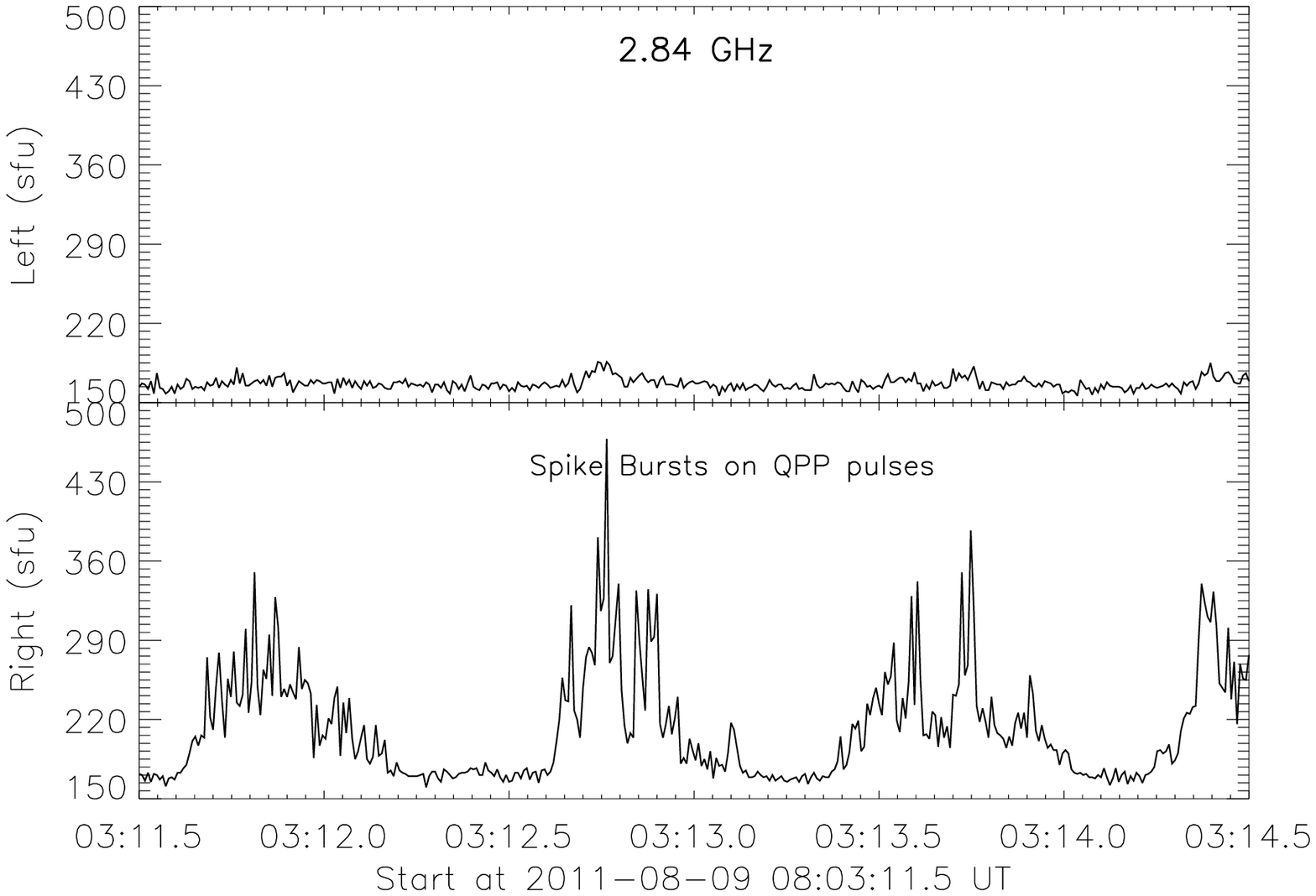}
  \includegraphics[width=7.5 cm]{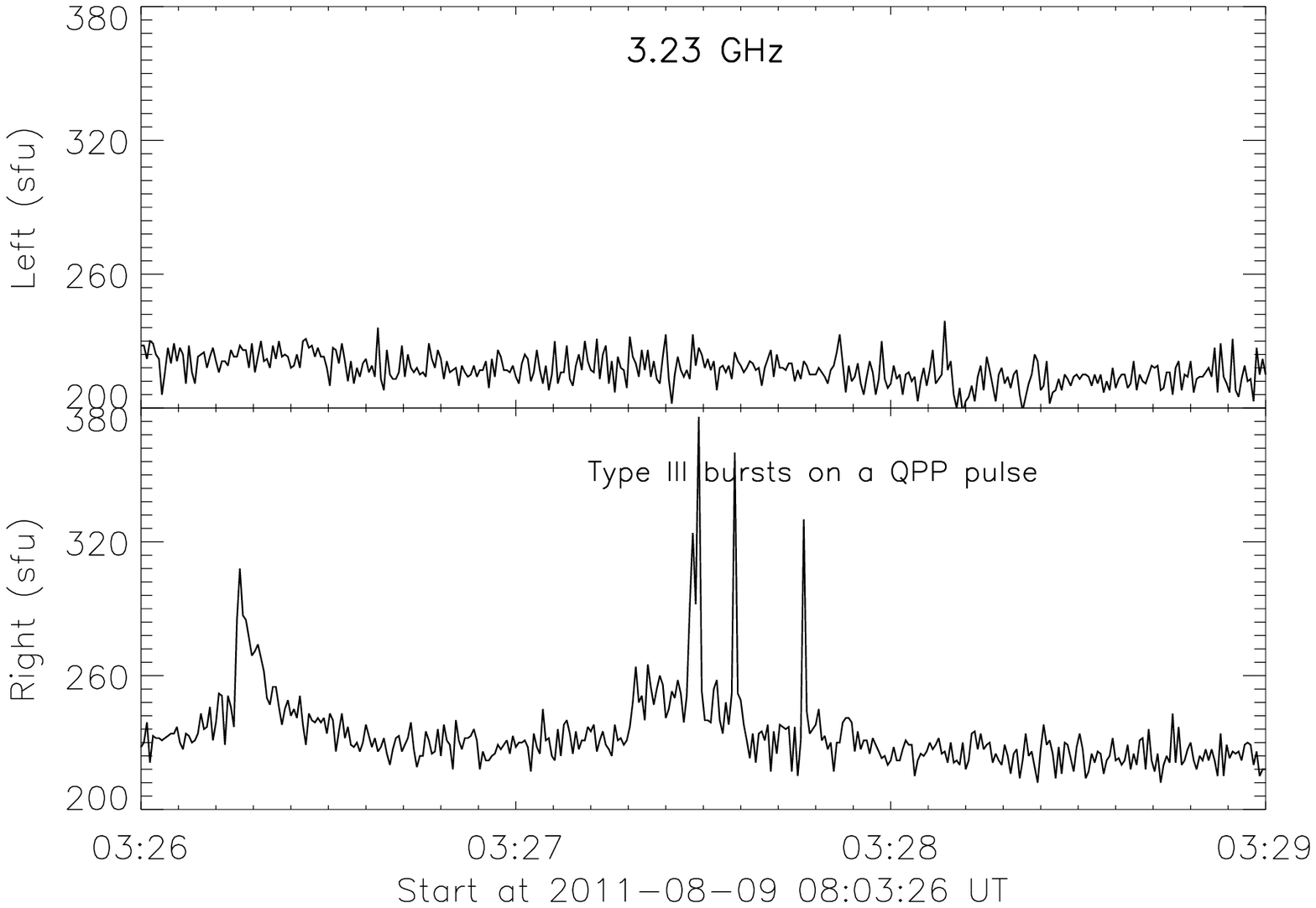}
  \caption{The top-left panel indicates the example method of calculating the frequency drifting rates
  of QPP pulses; and the top-right panel presents the distribution of the frequency drifting rates of QPP pulses.
  The bottom-left panel is the expanded profile of the QPP at a frequency of 2.84 GHz which presents details of the superfine spiky structures
  in several QPP pulses. The bottom-right panel is the expanded profile of a QPP pulse at a
  frequency of 3.23 GHz, which is structured with a group of type III bursts.}
\end{center}
\end{figure}

At the same time, the careful scrutinizing on each individual QPP
pulse presents another important feature: each pulse has a
superfine structures in shorter time scales. The expanded
spectrograms (the bottom panels of Figure 3) present the details
of the millisecond superfine structures in each QPP pulses. Here
we find that almost all of the microwave QPP pulses are structured
with clusters of millisecond spike bursts. The bandwidth of the
spike bursts is in the range of 20 - 60 MHz, the duration is in
the range of 8 - 24 ms (however, being restricted by the 8 ms
cadence of the telescope, we can not know if there are any spike
bursts with duration shorter than 8 ms. In fact, some spike bursts
are only observed by a single data point, this evidence is enough
to let us believe that some spike bursts may have a life-time
shorter than 8 ms), the flux intensity is in the range of 50 - 100
sfu, and the polarization degree is closed to 100\% (bottom-left
panel of Figure 4). Additionally, part of the millisecond spikes
show the obvious evidence of fast negative frequency drifting rate
with value of from -18 GHz s$^{-1}$ to -25 GHz s$^{-1}$, and the
average is -20 GHz s$^{-1}$.

The bottom-right panel of Figure 4 is the expanded profile at a
frequency of 3.23 GHz of a QPP pulse which shows that the QPP
pulse is structured with a cluster of type III bursts. The type
III bursts form a drifting pulsating structure (DPS) with local
GFDR of about 2.91 GHz s$^{-1}$, the pulsating period of about
0.16 s. As for each individual type III burst, the instantaneous
duration is about 8 - 24 ms, frequency bandwidth is about 500 -
1000 MHz, the polarization degree is above 85\%, the frequency
drift rate is in the range of from -14.58 GHz s$^{-1}$ to -20 GHz
s$^{-1}$, and the average value is about -17.30 GHz s$^{-1}$.

From the bottom-right panel of Figure 3, several segments of
strong zebra pattern (ZP) structures can be found at frequency of
2.60 - 2.75 GHz at 08:03:24 - 08:03:27 UT accompanying with the
microwave QPP pulses. Three zebra stripes can be seen in the
extended spectrogram in Figure 5 around 08:03:26 UT, the flux
intensities on individual stripe are degressive from the low
frequency to high frequency (about 1100 - 1300 sfu on the low
frequency stripe, 700 - 1000 sfu on the middle stripe, and 200 -
600 sfu on the high frequency stripe above the adjacent background
emission). It is strong right-handed circular polarization. The
frequency separation between the adjacent zebra stripes is about
70 - 100 MHz. Generally, the ZP structure is always regarded as
one of the most important microwave fine structure which can be
used to diagnose the magnetic field strength in the coronal source
regions (Zlotnik 2009, Chernov 2010). Adopting the similar method
of magnetic field estimation from ZP structure with double plasma
resonance model in the same frequency range of the work of Tan et
al. (2012), we may obtain the magnetic field strength in the
coronal source region of the ZP structure is about 147 - 210 G,
and the averaged value is about 178 G.

\begin{figure}         
\begin{center}
 \includegraphics[width=8.0 cm, height=9.0 cm]{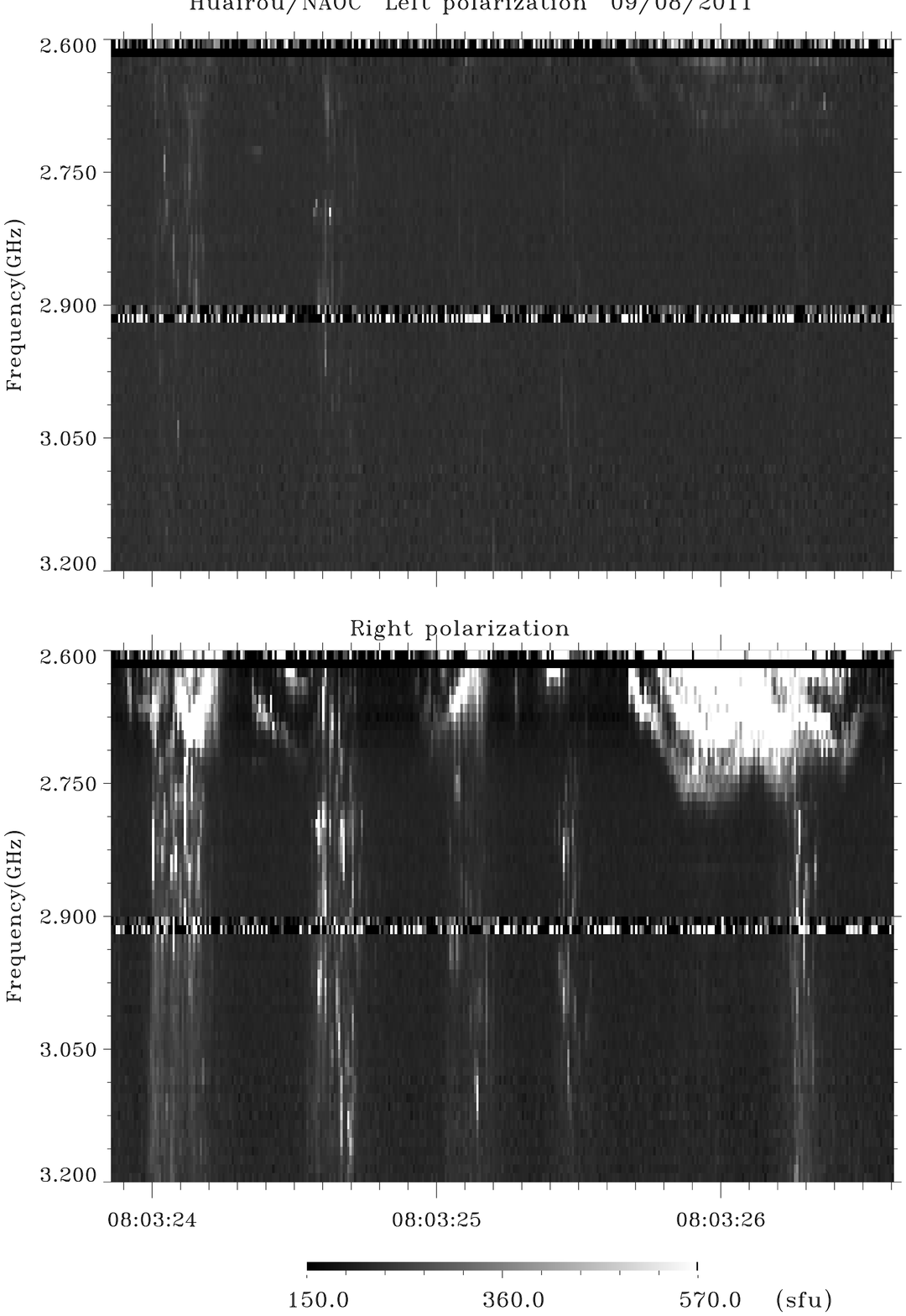}
  \caption{The spectrogram of a microwave zebra pattern structure occurred at 08:03:24 - 08:03:27 UT, accompanying with the microwave QPP at frequency of 2.60 - 2.75 GHz.}
\end{center}
\end{figure}

In brief, the RHESSI HXR observation at energy of 25 - 50 keV
implies a HXR QPP existence with period of 2.46 s. At the same
time, the microwave observation at frequency of 2.60 - 3.80 GHz
indicates the occurrence of a much fast QPP with average period of
0.705s. Each pulse of the microwave QPP is structured with
clusters of spikes or type III bursts in millisecond timescales.
All of them are strongly right-handed circular polarization; and
there are three distinct different frequency drifting rates:

(1) GFDR of microwave QPP pulse group: in several decades of MHz
s$^{-1}$, positive or negative;

(2) SPFDR of microwave QPP pulse: in several or more than 10 GHz
s$^{-1}$, positive or negative;

(3) SPFDR of millisecobd spike bursts or type III bursts: around
20 GHz s$^{-1}$, negative.

\section{Physical Discussions}

What physical implications hinted in the millisecond superfine
structures of each individual microwave QPP pulse? and how do we
understand the three distinct different kinds of frequency drift
rates?

At first, we need to determine the mechanism of the QPP. Here we
are only interested in the HXR QPP with period of 2 - 3 s and the
microwave VSP with period of subsecond. It is well known that
propagating MHD oscillation modes or the standing fast sausage
mode in dense magnetic traps may produce the QPP with period of
the above values (Roberts et al. 1984, Aschwanden 1987, Tan et al.
2010). Zaitsev et al. (1984) demonstrated that the variation of
plasma density in a magnetic loop by fast magnetosonic waves
excited by energetic protons in Cherenkov or bounce resonance
might produce QPP with period of seconds. Zaitsev et al. (1998,
2000) proposed another model of which a current-carrying plasma
loop should be a LRC-circuit resonator, which might cause periodic
modulation of the loop magnetic field, energy release rate, and
energetic electron production, therefore, the microwave emission,
and the corresponding period should be dependent on the
longitudinal electric current
($P_{LRC}\sim\frac{10^{12}}{I_{\varphi}}$, $I_{\varphi}$ is the
longitudinal electric current in the loop). In order to produce
the VSP in this work, the longitudinal electric current in the
loop should be in the range of (1.16 - 2.38)$\times10^{12}$ A.
Such value is very popular in a general flaring plasma loop (Gary
\& Demoulin 1995, Tan 2007, etc.). It seems that the above models
are the possible candidates for the QPP with periods in this work.
However, these models seem to be a bit of difficult to explain the
frequency drifting rates and the superfine structures with
clusters of millisecond spike bursts or type III bursts. Here, we
need a model which can explain the properties including the rapid
QPP and the superfine spiky structures.

Kliem et al. (2000) and Karlicky et al. (2004) proposed that
microwave QPP can be caused by quasi-periodic particle
acceleration from a highly dynamic regime of magnetic reconnection
in an extended large-scale current sheet above the flaring loop.
The reconnection is dominated by repeated formation and subsequent
coalescence of magnetic islands, known as secondary tearing modes.
This model can explain the QPP with period of 0.5 - 5 s in solar
flares. In particular, they may explain the global frequency
drifting rate of QPP as a motion of the plasmoid in the density
gradient of the solar atmosphere. Here, particles are accelerated
near the magnetic X-points in the DC electric field associated
with magnetic reconnection. The strongest electric fields occur at
the main magnetic X-points adjacent to the plasmoid, and a large
fraction of the accelerated particles may only be temporarily
trapped in the plasmoid; the accelerated process itself may form
an anisotropic velocity distribution, which may excite the
microwave emission. Based on particle-in-cell simulation, Karlicky
\& Barta (2007) found that electrons are accelerated most
efficiently in the region near the X-point of the magnetic
configuration at the end of the tearing process and the beginning
of plasmoid coalescence. The most energetic electrons are
localized along mainly the X-lines of the magnetic configuration.
During these process, plasma emission generated and propagated.

Tan et al. (2007) proposed that modulations of the resistive
tearing-mode (TM) oscillations in electric current-carrying
flaring plasma loops can also produce VSP with sub-second periods.
The duration of the VSP is dominated by the current density
distribution, magnetic configuration and plasma resistivity, as
well as the period of VSP is dominated by the total electric
current, the loop's geometrical parameters, and the distribution
of current density in the cross-section of the loop. The pulsating
emission is localized in some regions with small size, for
example, localized around magnetic islands in flaring plasma
loops. From the bandwidth of the emission we may estimate the
perturbation of the plasma density. The frequency drifting rate
can reflect the motion of the plasma loop and the energetic
particles. Combining all these observable parameters, we may probe
many physical conditions and their evolutions.

As for the mechanism of the millisecond spike bursts or
narrow-band type III bursts, Benz (1986) proposed that decimeter
narrow band millisecond spikes are coherent emission as a
signature of electron acceleration in flare, and the flare energy
release must be fragmented with each spike indicating a single
energy release episode in the smallest time and space scale
originated from magnetic reconnection sites (Huang \& Nakajima
2005, Wang et al. 2008). Fleishman \& Melnikov (1999) made a
detailed comparison of the observed spike properties with various
theoretical models, and found that the electron cyclotron maser
emission (ECME) driven by nonthermal electrons can explain almost
all the observed properties. Fleishman et al. (2003) proposed that
the source of spike cluster should be a coronal loop filled by
fast electrons and relatively tenuous background plasma. Each
spike is generated by ECME in a local small source inside the
loop. ECME takes place only when the following conditions
satisfied (Melrose \& Dulk 1982):
\begin{equation}
\omega-\frac{s\omega_{ce}}{\gamma}-k_{\parallel}v_{\parallel}=0
\end{equation}
and
\begin{equation}
\omega_{ce}>>\omega_{pe}
\end{equation}
Here, $\omega$ is the emission frequency, $\omega_{ce}$ is the
electron gyro-frequency, $\omega_{pe}$ is the electron plasma
frequency, $s$ is the harmonic number, $\gamma$ is the Lorentz
factor of the energetic electrons, $k_{\parallel}$ and
$v_{\parallel}$ are the parallel components of the wave number and
particle velocity, respectively. Equation 3 implies that the
source region of ECME must have a relatively strong magnetic
field. Because of the accompanying of ZP and microwave QPP
structures, we may suppose that their source regions are close to
each other, and the magnetic field strengths are also similar to
each other. However, the estimation of ZP structures indicates
that the magnetic field strength is about 147 - 210 G, and the
corresponding gyrofrequency $f_{ce}$ is about 412 - 588 MHz, which
is much lower than the observed frequency ($\sim$ 3.00 GHz). This
fact implies that ECME seems not to be the formation mechanism of
the millisecond spiky bursts observed in this work.

It is well known that there is another kind of coherent emission:
plasma emission, which is always generated from the coupling of
two excited plasma waves with frequency of about 2$f_{pe}$ and
weakly polarizations, or the coupling of an excited plasma wave
and a low-frequency electrostatic wave with frequency of about
$f_{pe}$ and strongly polarizations (Zheleznyakov \& Zlotnik 1975,
Chernov et al. 2003). Generally, the plasma emission is triggered
by some Langmuir turbulence produced from nonthermal energetic
electrons.

Combining the above analysis, we may plot qualitatively the
physical processes associated with the QPP as in Figure 6:

\begin{figure}         
\begin{center}
 \includegraphics[width=7.5 cm]{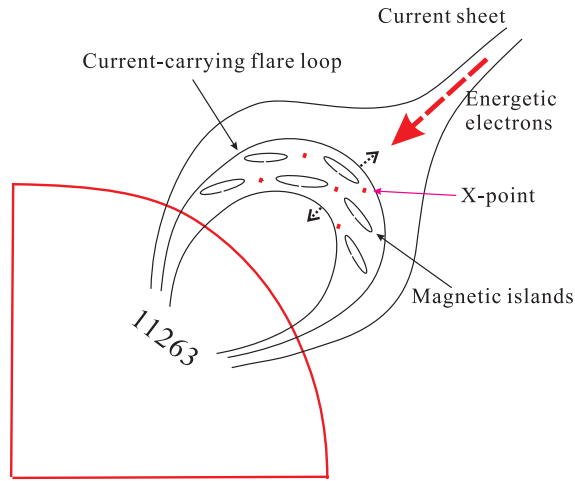}
  \caption{A schematic cartoon which shows the processes of electron acceleration above and in the flare loop, propagation, the formation and the distribution of magnetic islands
  in the current-carrying flare loop.}
\end{center}
\end{figure}

(1) From the idea of Kliem et al. (2000), we may propose that the
energetic electrons accelerating from a highly dynamic magnetic
reconnection in an extended large-scale current sheet above the
flaring loop propagate downwards, interact with the plasma in the
flaring loop, and produce HXR bursts by bremsstrahlung mechanism.
The TM oscillation in the current sheet may modulate the HXR
emission and form the HXR QPP (the big red arrow in Figure 6);

(2) The energetic electrons propagating downwards may impact on
the flaring loop and trigger Langmuir turbulence and plasma waves
in the loop. The coupling between the plasma wave and the low
frequency electrostatic wave emits the microwave radiation by the
mechanism of plasma emission. From the idea of Tan et al. (2007),
the modulation of the TM oscillation in the current-carrying
plasma loop may lead to the formation of microwave QPP. As the
frequency of plasma emission depends on the plasma density, the
motion of the flaring loop with respect to the background may
change the plasma density, and result in the frequency drifting
rate of the microwave QPP pulse groups. The motion of the
energetic electron beam with respect to the background result in
the frequency drifting rate of the QPP pulses. From the mechanism
of plasma emission, the GFDR of QPP pulse group is due to the
motion of the plasma loop (Aschwanden \& Benz 1986):
\begin{equation}
\frac{\partial \omega}{\partial
t}=-\frac{\omega}{2H_{n}}v_{loop}\cos \theta
\end{equation}
Here, $H_{n}$ is the scale length of the density inhomogeneity in
the ambient plasma of the loop, $v_{loop}$ is the velocity of the
loop's motion, and $\theta$ is the angle between the line-of-sight
and magnetic field line. Generally, we may assume $H_{n}\sim
10^{4}$ km, and $\theta\sim0$. Substitute the GFDR of QPP group
into Equation (4), we may obtain the velocity of the loop is about
550 km s$^{-1}$ upwards in group A, 526 km s$^{-1}$ downwards in
group B, and from 655 km s$^{-1}$ downwards to 863 km s$^{-1}$
upwards in group C.

In Equation (4), if we replace the $v_{loop}$ by the velocity of
the energetic electron beams ($v_{beam}$), then we may obtain the
SPFDR of QPP pulse. Substitute the values of SPFDR of QPP pulses
into Equation (4), we obtain the velocity of the energetic
electron beams as about $1.94\times10^{4}-1.75\times10^{5}$ km
s$^{-1}$. And the corresponding kinetic energy of the energetic
electrons is about 1.07 - 86.7 keV.

(3) The TM instability cause the formation of many magnetic
islands in the current-carrying plasma loop. Each X-point between
the adjacent magnetic islands will be a small reconnection site
and make a secondary acceleration on the ambient electrons (Drake
et al. 2006). These accelerated electrons impact on the adjacent
plasmas around the X-point, trigger the Langmuir turbulence and
plasma waves, produce microwave bursts by plasma mechanism. Such
microwave bursts are just the spikes or type III bursts. As there
are many magnetic X-points in the current-carrying plasma loop,
each X-point will be a small reconnection site, and the region
around each X-point will be a source of a spike burst or type III
burst. As a result, the spike bursts or the type III bursts can
arise in huge clusters. Because from X-point to the space between
two adjacent rational surfaces the plasma density always has a
negative gradient (Furth et al 1973, Drake et al. 2006), so the
frequency drifting rates of spike bursts or type III bursts are
negative. Because the scale length of the density inhomogeneity
($H_{n}$) around the X-point is shorter than that in the ambient
plasma of the loop, from Equation (4) we may conclude that the
frequency drifting rates of spike bursts or type III bursts are
higher than that of the microwave QPP pulses.

\section{Conclusions}

Based on the above observations and physical analysis, we may come
to the following conclusions:

(1) Just before the flare maximum, there are HXR QPP and microwave
QPP with durations of about 20 s. The HXR QPP belongs to a SPP,
and the microwave QPP belongs to a VSP. Each individual microwave
QPP pulse is structured with clusters of millisecond bursts, most
of the millisecond bursts are cluster of spike bursts, and the
rest are type III bursts. There are three distinct different
frequency drifting rates: GFDR of microwave QPP pulse group, SPFDR
of microwave QPP pulse, SPFDR of millisecond spike bursts or
narrow band type III bursts.

(2) The physical processes associated with the QPP can be
described as: the energetic electrons accelerating from a highly
dynamic magnetic reconnection in an extended large-scale current
sheet above the flaring loop propagate downwards, interact with
the plasma of the flaring loop, and produce HXR bursts by
bremsstrahlung mechanism. The TM oscillation in the current sheet
may modulate the HXR emission and form HXR QPP; the energetic
electrons propagating downwards may produce Langmuir turbulence
and plasma waves in the loop, trigger the plasma emission. The
modulation of the TM oscillations in the current-carrying plasma
loop may lead to formation of microwave QPP. The TM instability
cause the formation of magnetic islands in the current-carrying
plasma loop. Each magnetic X-point between the adjacent two
magnetic islands will be a small reconnection site and make a
secondary acceleration on the ambient electrons. These accelerated
electrons may impact on the ambient plasma and trigger the
millisecond spike clusters or the group of type III burst by
plasma mechanism.

(3) Each magnetic X-point is an energy releasing site in the
current-carrying plasma loop, and possibly each millisecond spike
burst or type III burst is one of the elementary burst (EB). Large
numbers of EB clusters form an intense flaring microwave burst.

(4) We may apply the above conclusions to analyze the conditions
of the source regions, e.g. the GFDR of microwave QPP pulse group
can be adopted to estimate the motion of the flaring loop, the
SPFDR of microwave QPP pulse can be adopted to estimate the
velocity of the energetic electrons and the corresponding
energies, and the SPFDR of millisecond spike bursts or type III
bursts can be applied to estimate the density inhomogeneity
($H_{n}$) around the X-point in the flaring plasma loops.

However, as we are lack of imaging observations with spatial
resolutions in the corresponding frequency range, there are many
unresolved problems of the QPPs, for example, the spatial
behaviors, the spatial scales of the source region, etc. To
overcome these problems, some new instruments are needed, for
example, the constructing Chinese Spectral Radioheliograph (CSRH,
0.4 - 15 GHz) in the decimetric to centimeter-wave range (Yan et
al, 2009) and the proposed American Frequency Agile Solar
Radiotelescope (FASR, 50 MHz - 20 GHz) (Bastian, 2003). Maybe,
when these instruments begin to work, we will get more and more
cognitions of the solar activities.

\acknowledgments

The authors would like to thank the referee for the helpful and
valuable comments on this paper. Thanks are also due to the GOES,
RHESSI, SDO/AIA, and SBRS/Huairou teams for the observational
data. This work is mainly supported by NSFC Grant No. 11103044,
10921303, MOST Grant No. 2011CB811401, the National Major
Scientific Equipment R\&D Project ZDYZ2009-3.

\end{document}